\documentclass[useAMS]{gGAF2e}
\usepackage{graphicx}
\usepackage{amsmath}
\newcommand{\pa}{\partial}

\usepackage[percent]{overpic}
\usepackage{xcolor}
\usepackage{pict2e}
\usepackage{pbox}
\usepackage{amsmath,amssymb}
\usepackage{booktabs}
\usepackage{multirow}

\newcommand{\EQ}{\begin{equation}}
\newcommand{\EN}{\end{equation}}
\newcommand{\EQA}{\begin{eqnarray}}
\newcommand{\ENA}{\end{eqnarray}}

\newcommand{\JJ}{\mbox{\boldmath $J$} {}}
\newcommand{\UU}{\mbox{\boldmath $U$} {}}

\newcommand{\De}      {\mathrm{D}}
\renewcommand{\vec}[1]{\mbox{\boldmath{$#1$}}}

\newcommand{\Div}     {\vec{{\bm \nabla}}{\bm \cdot}}
\newcommand{\nab}     {\nabla}
\newcommand{\Laplace} {\nabla^2}

\newcommand{\Strain}{\mbox{\boldmath ${\sf S}$} {}}

\newcommand{\Av}      {\vec{A}}
\newcommand{\Bv}      {\vec{B}}
\newcommand{\Ev}      {\vec{E}}

\begin{document}
\doi{}
\issn{} \issnp{} \jvol{00} \jnum{00} \jyear{2019} 
\markboth{Chatterjee}{Alfv\'en waves in solar atmosphere}

\title{Testing Alfv\'en wave propagation in a "realistic" set-up of the solar atmosphere}
\author{Piyali Chatterjee, \\ Indian Institute of Astrophysics, \\ II Block, Koramangala, \\ Bengaluru-560034, India}

\maketitle

\begin{abstract}
We present a radiative magneto-hydrodynamic simulation set-up using the {\sc pencil code} to study the generation, propagation and dissipation of Alfv\'en waves in the solar atmosphere which includes a convective layer, photosphere below and chromosphere, transition region and the corona above.  We prepare a setup of steady-state solar convection where the imposed external magnetic field also has reached the final value gradually starting from a very small value. From that state, we start several simulations by varying the magnetic Prandtl number and the forcing strengths. 
We find the propagation characteristics of waves excited in this simulation run depend strongly on the magnetic Prandtl number and the wave number of the forcing. For magnetic Prandtl number of unity, we obtain localized heating in the corona due to shock dissipation. 
\end{abstract}


\section{Introduction}
The reason for a hot solar corona still remains a grand challenge problem in the field of plasma physics.  The observation of highly ionized Iron in the solar corona requires evenly high temperatures to maintain these ionization equilibria \citep{Grotrian34}. A high degree of ionization of several elements in the corona also points to high temperatures \citep{Edlen43}. A canonical value of a million Kelvin (MK) usually quoted for coronal plasma temperatures certainly cannot be the black body radiation temperature as the plasma is tenuous and optically thin. It may represent the kinetic electron temperature, $T_e$ - which has been measured using line ratios of coronal emission lines like Si XII/Mg X, Si XII/Mg IX, or Mg X/Mg IX observed with the CDS instrument on SOHO and assuming that the plasma is isothermal and has a Maxwellian velocity distribution  giving values like 1.6 MK in coronal streamers and 0.8 MK in polar coronal holes \citep{Fludra_etal99} or from radio emission due to free-free thermal Bremsstrahlung  around optically thick 170 MHz emission giving temperatures of about 0.64 MK (Fokker 1966).  It may also be the ion temperature ($T_i$) measured using the line widths of emission lines and can be significantly hotter than $T_e$ \citep{LC09} or the proton temperature ($T_p$) the upper limit of which could be $\sim 6$ MK, obtained from the HI Ly $\alpha$ line widths in polar coronal holes \citep{Kohl_etal98, Vasq_etal03}. 
Also, both $T_i$ and $T_p$ are said to be highly anisotropic depending on if the measurement is parallel or 
perpendicular to the radial magnetic field. Therefore, it is not clear to us what the temperature, $T$ used in the 
magneto hydrodynamic (MHD) approximation for the coronal plasma (described with a single fluid density, $\rho$) should correspond to out of $T_e$, $T_p$ and $T_i$. Nevertheless, the 
problem remains to devise one or several mechanisms to accelerate the electrons, protons and ions in the plasma to high enough energies to be able to explain the emission line ratios, the line widths and the radio emission from the solar corona.   
Two competing theories acknowledged by experts are the wave heating or AC models and the DC models which include nanoflares \citep{Parker81, Parker83}, heating along separatrix layers in the corona due to {\em flux tube tectonics} \citep{Priest02} and slow diffusive \citep{BBP13} and fast intermittent reconnections \citep{BP11, BP13}. Many early authors introduced the idea that Alfv\'en wave steepening and dissipation in the solar atmosphere may be  a candidate for the million Kelvin corona \citep{A47,Oster61, FP58}. 
Alfv\'en or shear Alfv\'en waves are incompressible normal modes of magneto hydrodynamic  equations charecterized by transverse motion of magnetic field lines. A peculiar property of these incompressible waves is that they are exact solutions of the full non-linear MHD equations irrespective of their amplitudes and thus can travel long distances 
into the heliosphere before depositing their energy as compared to compressive modes \citep{Ofman02}. 
The other solutions of MHD 
equations -slow magneto acoustic mode will decay before even reaching the corona and can at most 
contribute to heating the  upper chromosphere, the fast magneto acoustic mode will likely undergo refractions in a highly stratified atmosphere before reaching the corona. Solar chromosphere is believed to be a vast reservoir of wave energy --compressible or incompressible modes -- that has been observed \citep{Morton_etal12} using the ROSA imager for the Dunn solar telescope. Even though incompressible Alfv\'en waves can travel far into the corona without refraction, these modes are very difficult to dissipate under solar atmospheric conditions and require large Alfv\'en speed gradients or presence of highly non uniform magnetic fields like flux tubes. Observationally Alfv\'en waves have also been observed in coronal holes and in the fast solar wind  as observed by the Helios and Ulysses missions \citep{Marsch91, Gold95}. It is therefore possible that Alfv\'en wave heating takes place in coronal holes at heights above where the fast solar wind originates. We believe that in the lower corona, or in the quiet sun and above active regions, the Ohmic heating due to magnetic reconnection can compete with Alfv\'en wave dissipation. 

Some authors, using numerical MHD models, but extending only up to $\leq 37$ Mm above the photosphere, 
report that Parker's nanoflare model is sufficient to heat the corona to million degree Kelvin \citep{Gud02, Gud05, BP11, BP13}. These state-of-the-art three dimensional Cartesian models are also referred to as ``realistic" because of their 
treatment of the magnetic field aligned thermal conduction, radiative transfer assuming local 
thermodynamic equilibrium or LTE  \citep{BP11, BBP13, Chen_etal14, Rempel17} or  the 
treatment of partial ionization and non-LTE effects in the 
chromosphere (Gudiksen et al. 2011). Recently, \cite{RH17} have used the {\sc Flash} code as well as the 
{\sc Pencil code} \citep{BD02} to study Alfv\'en wave dissipation in a two dimensional 
solar atmosphere between $1~\mathrm{Mm}<z<10$ Mm for a simulation duration of $\sim 500$ sec solar time. \cite{KhoCally11} have also used a 2.5 D MHD model up to 1 Mm height to study the conversion of fast magneto acoustic modes to Alfv\'en modes in the chromosphere in the presence of an inclined magnetic field of strength~500 G. \cite{vB_etal17} used a 3D MHD extending up to a height of 50 Mm but with reduced physics to study 
AC heating in the form of Alfv\'en wave turbulence and showed that it is sufficient to heat coronal loops to MK temperatures.  
But, to the best of our knowledge, the two models of heating (AC and DC) have never been compared using the same numerical code and under similar thermodynamic conditions.  This work has been initiated with the aim of building such models in 2D, 2.5D and 3D where the contribution from the MHD shock waves, Alfv\'en wave dissipation, reconnection and nano flare heating can be studied.  
In the present work, we would like to clarify that whenever we use term  ``realistic", we refer to the vertical stratification of the atmosphere as well as to the fact that we use field aligned anisotropic thermal conduction and (LTE) radiative transfer.  The convection developing in our 2D model is far from the real Sun since 3D Rayleigh-B\'enard hexagonal convection cells are fundamentally 
different from the 2D convection rolls. Also, the direction of energy cascade in wavenumber space is opposite in 2D and 3D. Previous authors \citep{BP11, BP13, BBP13} have already 
used the {\sc pencil code} to model a ``realistic" solar  corona. For example, see the sample set up -- \texttt{solar-atmosphere-magnetic} -- for a set-up spanning $-2.6~\mathrm{Mm}< z<0.6$ Mm, where $z=0$ denotes the position of the photosphere and, another sample set-up -- \texttt{corona}  -- spanning $0<z<40$ Mm.  
In the latter set up the solar atmosphere is driven at the lower boundary by a granulation driver which mimics the solar photospheric convection and a magnetogram driver which provides a time varying vertical
magnetic field at the photosphere.  The reader is also referred to the manual which comes with the 
{\sc pencil code} distribution for further details of the equations.  In this paper, we present the creation 
of a ``realistic" but two dimensional solar atmosphere 
including subsurface convection, photosphere, chromosphere and a corona as well as for the first time incorporate the semi-relativistic Boris correction into the {\sc Pencil code}. Boris correction, which is a way of reducing numerical diffusion and the need for using large explicit diffusion in MHD simulations, solves the semi-relativistic magnetohydrodynamics equations with an artificially reduced speed of light. This correction was introduced by \cite{Boris70} and later incorporated in MHD codes like 
BATS-R-US \citep{Gombosi_etal02} and, MuRAM \citep{Rempel17}. An alternate fully implicit time stepping scheme - Implicit Scheme with Limited Numerical Dissipation (ISLND)-was developed and benchmarked with Boris correction by \cite{Toth_etal11} with comparable results. We also use a specific form of the hyperbolic heat transport equation into the {\sc pencil code}.  We then test our model set up for the propagation and dissipation of MHD waves in the presence of an external oblique magnetic field.

\section{Setting up the {\sc pencil code} for the solar atmosphere}
The computational domain consists of a box, the horizontal extents of which are -6.25 $  \mathrm{Mm} < x < $6.25  $\mathrm{Mm}$, and the vertical extent is -10 $\mathrm{ Mm}< z <$ 15.0 $\mathrm{ Mm}$. A constant gravity, $g_z$, with magnitude $-2.74\times10^4$  cm s$^{-2}$ points in the negative $z$-direction. The $z =0$ height denotes the solar photosphere and the region between $-10$ Mm $<z<0$ represents uppermost part of the solar convection zone.   
The box is resolved using a uniformly spaced grid with $dz, dx=48$ km. We use the fully compressible higher-order 
finite difference tool, the {\sc Pencil Code}\footnote{{https://github.com/pencil-code/}} for these calculations. We use a sixth order finite difference scheme and a second order Runge-Kutta time stepping scheme out of several other options available with the {\sc pencil code}. This  code is highly modular and can easily be adapted to different types of computational MHD problems. 
We solve the following set of compressible MHD equations.
The continuity equation is given by
\begin{equation}
\label{eq:continuity}
\frac{\De\ln \rho}{\De t}
  = - \Div\UU,
  \end{equation}
    
where
$\De/\De t \equiv \partial/\partial t + \UU{\bm \cdot}\nab$ denotes the Lagrangian derivative with respect to the local 
velocity of the gas, $\UU$, and $\rho$ is the local plasma density in units of the photospheric density, $\rho_0=2.7\times10^{-7}$ g cm$^{-3}$. 

{\em Equation of state:} We use an equation of state which includes ionization calculated assuming {\em{local thermal equilibrium}}(LTE). The pressure, $p$, is given by, $$p=\frac{\rho R_g T}{\mu(T)}.$$
where, $R_g=k_B/m_u$ is the ideal gas constant, $T$, is the temperature and $\mu$, is the effective mass given by $$\mu=\frac{4 x_\mathrm{He}+1}{y_\mathrm{H}(T)+x_\mathrm{He}+1}.$$ We consider the number fraction of Helium, $x_\mathrm{He}$, to have a constant value of 0.089, whereas the fraction of ionized Hydrogen, denoted by $y_\mathrm{H}$ is a function of temperature. To calculate, $y_\mathrm{H}(T)$, at each time step,  using the Saha's ionization formula. We solve the following quadratic equation in $y_\mathrm{H}$ \citep{BhB16},
\EQ
\frac{y_\mathrm{H}^2}{1-y_\mathrm{H}}=q,
\EN
with the equilibrium reaction constant, $$q=\frac{\rho_\mathrm{e}}{\rho}\left(\frac{\chi_\mathrm{H}}{k_B T}\right)^{-3/2}\exp(-\chi_H/k_B T) ,$$ and whose solution is given by,
\EQ
\label{eq:saha}
y_\mathrm{H} =  \frac{2\sqrt{q}}{\sqrt{q}+\sqrt{4+q}},
\EN
with, 
\EQ
\nonumber
\rho_e = (1+4x_\mathrm{He}) m_u \left(\frac{m_e \chi_\mathrm{H}}{2\pi \hbar^2}\right)^{3/2}.
\EN

$\chi_\mathrm{H}=13.6$ eV, $m_u$ is 1 amu and, $m_e$, the mass of electron. 
The momentum equation is,

\EQ
  \begin{aligned}
    \frac{\De\UU}{\De t}
   =&-\frac{{\bm \nabla} p}{\rho}   
      + g_z \vec{\hat{z}}                         
      + \frac{\JJ\times\Bv}{\rho}+\vec{F}^\mathrm{corr}_L  \\
      &+ \rho^{-1}\mathbf{F}_\mathrm{visc},
\end{aligned} \label{eq:NS}
\EN
where $\JJ$ is the current density, $\Bv$ is the magnetic field, $\vec{F}^\mathrm{corr}_L$ is the semi-relativistic correction due to Boris (1970) which has been discussed in \S\ref{sec:boris}. The viscous force is modelled as,
\EQ
\begin{aligned}
\rho^{-1}\mathbf{F}_\mathrm{visc} = &\nu \left( \Laplace\UU + \frac{1}{3}{\bm \nabla}\Div\UU
      + 2\Strain {\bm \cdot}{\bm \nabla}\ln\rho\right) \\
      &+\zeta_\mathrm{shock}\left[{\bm \nabla} \left({\bm \nabla}\cdot\UU\right)+\left({\bm \nabla}\ln\rho+{\bm \nabla}\ln\zeta_\mathrm{shock}\right){\bm \nabla}\cdot\UU \right],
      \end{aligned}
      \label{eq:visc}
\EN
where, $\nu$ is the height-dependent kinematic viscosity, and $\Strain$ is the traceless rate-of-strain tensor. We also use an enhanced viscous force at the shock fronts. 
The expression for the coefficient $\zeta_\mathrm{shock}$ in equation~\ref{eq:visc} is given by,
\EQ
\zeta_\mathrm{shock}=\nu_\mathrm{shock} \langle \mathrm{Max}_3\left[(-{\bm \nabla}{\bm \cdot}\UU)_{+}\right]\rangle,
\EN
where, $\langle \mathrm{Max}_3\left[(-{\bm \nabla}{\bm \cdot}\UU)_{+}\right]\rangle$ means that to each grid point, we assign a value given by the maximum of the positive flow convergence ($-{\bm \nabla}{\bm \cdot}\UU > 0$) over three neighboring grid points along each spatial dimension and then smooth the result using a running mean over three neighboring grid points along each coordinate direction. 
The induction equation is solved for the magnetic vector potential, $\Av$, using the uncurled induction equation,  
\begin{equation}
\label{eq:induc}
  \frac{\partial\Av}{\partial t}
  = \UU\times\Bv - \eta\mu_0\JJ +{\bm \nabla}\varPsi.
\end{equation}

In presence of an external magnetic field, $\Bv_\mathrm{ext}$, $\Bv=\Bv_\mathrm{ext}+\Bv'$ and ${\bm \nabla}\times\vec{A} = \Bv'$ and $\eta$ denotes molecular magnetic diffusivity. Gauge freedom allows us to set $\varPsi=0$ (Weyl gauge) at all times. 
 
The initial stratification of temperature is obtained by collating the Model S \cite{Christensen_etal96} for the interior and the atmospheric model by \cite{Verna_etal81}. The initial density stratification corresponding to this temperature is obtained by solving the hydrostatic balance subjected to the ionized ideal gas equation of state with ionization fraction given by the  Saha-ionization formula (see equation~\ref{eq:saha}). 
 
Finally, we have for the temperature equation, with turbulent diffusion, $\chi_t$,

\begin{eqnarray}
\label{eq:entropy}
\nonumber
   \rho c_V T\frac{\De \ln T}{\De t}
   =  -(\gamma-1)\rho c_v T \Div \UU+\Div(\vec{q}_\mathrm{cond}+\vec{q}_\mathrm{rad}) + \Div(\rho T \chi_t {\bm \nabla} \ln T)\\
      + \eta\mu_0\JJ^2
      + 2\rho\nu{\sf{S}}_{ij}^2 +{\mathbf \rho\zeta_\mathrm{shock}\left({\bm \nabla}{\bm \cdot}\UU\right)^2}
      -\rho^2\varLambda(T) .
\end{eqnarray}

The Spitzer heat conduction flux is denoted $\vec{q}_\mathrm{cond}$ and described in \S \ref{sec:hyperbolic}, ${\sf{S}}^2_{ij}$ denote the square of the tensorial components of the rate-of-strain tensor, $\Strain$ summed over all indices, $i, j$ and, $c_V$ is the specific heat capacity at constant volume. 

{\em Radiative transport:} The radiative flux is denoted by $\vec{q}_\mathrm{rad}$ and is calculated using the method of long characteristics as described in \cite{H_etal06}. To compute $\Div\vec{q}_\mathrm{rad}$, we solve the equation of radiative transport by adopting a gray approximation, and by neglecting scattering contributions. 
\EQ
\label{eq:rad}
\hat{\vec{n}}{\bm \cdot}{\bm \nabla} I = -\kappa_\mathrm{tot}\rho{(I-S)},
\EN
where, $I(x,z,t,\hat{\vec{n}})$ is the specific intensity along direction $\hat{\vec{n}}$. The source function, $S=(\sigma_\mathrm{SB}/\pi)T^4$ is the frequency integrated Planck's function with $\sigma_\mathrm{SB}$ being the Stefan-Boltzman constant. The integration of equation~(\ref{eq:rad}) over solid angle, $\varOmega$, will give us $\Div\vec{q}_\mathrm{rad}$ by,
\EQ
\Div\vec{q}_\mathrm{rad}=\kappa_\mathrm{tot}\rho\oint_{4\pi}(I-S){\mathrm{d}}\varOmega .
\EN

For this angular quadrature we use eight rays, 4 along $x$ and $z$ axes and 4 along face diagonals of the $x-z$ grid, and correctly scale the angular weight factors for two dimensionality \citep{BB14}. We do not use tabulated opacities, rather use analytical power-law fits  to Rosseland mean opacity functions.  We use solar abundances X=0.7381, Y=0.2485, metallicity Z=0.0134. The bound-free, free-free and $\mathrm{H}^{-}$ opacities are combined to give the total opacity, $\kappa_\mathrm{tot}$.  The Rosseland mean of the bound-free and free-free opacities can be written in the Kramers power law form \citep{HK_book94},
$$\kappa_\mathrm{bf+ff}\sim 4\times 10^{25}Z(X+1)\rho T^{-7/2} \mathrm{cm}^2\mathrm{g}^{-1},$$
whereas, the $\mathrm{H}^-$ opacity which cannot be expressed in a Kramers power law form is given by,
$$\kappa_\mathrm{H^-}\sim 1.25\times10^{-29}Z\rho^{1/2} T^9 \mathrm{cm}^2\mathrm{g}^{-1}.$$
The conductive opacity due to the electron scattering, for partial ionization  is
$$\kappa_\mathrm{c}=2.6\times10^{-7}\frac{T^2}{\rho} \mathrm{cm}^2\mathrm{g}^{-1}.$$
The total opacity is then given by
$$\frac{1}{\kappa_\mathrm{tot}}=\frac{1}{\kappa_r}+\frac{1}{\kappa_c},$$
where,
$$\frac{1}{\kappa_r}=\frac{1}{\kappa_\mathrm{bf+ff}}+\frac{1}{\kappa_\mathrm{H^{-}}}.$$

Following the MuRAM code \citep{Rempel17} where the radiative module is shut off once transition region is reached, we smoothly put the source function, $S$ and the opacity, $\kappa_\mathrm{tot}$, to zero above $z = z_\mathrm{cutoff}=1.5$ Mm to avoid contribution from the transition region and corona in the downward directed rays. For a high temperature in the corona, the Planck source function used in the radiative transfer equation will have a huge value and therefore is not realistic. Hence we do not take such back reaction into account. Additionally, in order to limit the numerical value of radiative heating, we use an upper bound $10^3$ ergs cm$^{-3}$ s$^{-1}$ for $\Div\vec{q}_\mathrm{rad}$ before adding it to equation~(\ref{eq:entropy}). The transport of radiation through chromosphere normally gives rise to cooling ($\Div \vec{q}_\mathrm{rad} < 0$) rather than heating ($\Div \vec{q}_\mathrm{rad} > 0$) since some of the solar energy passing through the chromosphere is absorbed by atoms to make transitions to higher energy states (or even get ionized) and therefore is not available for raising the temperature of the chromospheric plasma. The situation is highly dynamic during first few mins (solar time) of the simulation run when shocks are passing repeatedly through this region. After 1 hour of solar time, when the convection has settled to a steady state, changing the upper limit to a much higher positive value does not affect the simulation. In fact, $q_\mathrm{rad}$ does not exceed 25\% of its maximum negative value in the domain. The last term in equation ~(\ref{eq:entropy}) is the optically thin radiative cooling which is activated in the solar corona only for $z > z_\mathrm{cutoff}$ and is obtained by best fits to \cite{Cook_etal89} by using a linear piecewise interpolation algorithm in $\log T$. This corresponds to setting the parameter \texttt{cool\_type=5} in the \texttt{solar\_corona} special module in the {\sc{Pencil code}}.

{\em Boundary conditions and explicit dissipation:} The lower boundary at $z_B$=-10 Mm is forced by a sinusoidal velocity forcing with an amplitude $u_f=(L_{\odot}/4\pi \rho(z_B) R_\odot^2)^{1/3}$  and the top boundary at $z_T$= 15.0 Mm is open.  The solar luminosity and radius are denoted by symbols $L_{\odot}$ and $R_{\odot}$, respectively. The purpose of the lower boundary driving is to excite waves of particular horizontal wavenumber comparable to the granulation scale. It also helps development of convection faster than cases where we do not include this driving. 
The density and temperature are considered to be in hydrostatic balance at either vertical boundary. This means that the slopes, $\mathrm{d}\ln\rho/\mathrm{d}z$ and $\mathrm{d}\ln T/\mathrm{d}z$, in the ghost zones are set to values expected from hydrostatic balance at the vertical boundaries. The $x$-boundaries are periodic. Our simulation set-up is different from that presented in \cite{Rempel17} in the way that we haven't used opacity tables and a tabulated equation of state. Analytical expressions for the opacity approximating the tabulated opacity tables and the Saha ionization formula has been used. The treatment of the top boundary is also different in that in \cite{Rempel17}, a sponge like top boundary is used for the fluid velocity in contrast to here where we put the vertical derivatives of $U_x$ and $U_y$ to zero and we also set $\mathrm{d}U_z/\mathrm{d}z=0$ only if we detect an outflow, else if we detect any plasma inflow at any horizontal grid point at the top boundary, we set $U_z=0$, and for the magnetic field, we replicate the value of the electromotive force, $\mathcal{E}=-\UU\times\Bv$ of the last grid point to all ghost cells. Our top boundary allows reflection of the MHD waves in spite of allowing matter and energy to flow out since we do not solve for the equation of characteristics at the boundary for MHD waves.  For specific intensity, we assume zero intensity at the top and set it equal to the source function at the bottom. 
\begin{figure}
\begin{overpic}[width=0.49\textwidth,height=.72\textwidth]{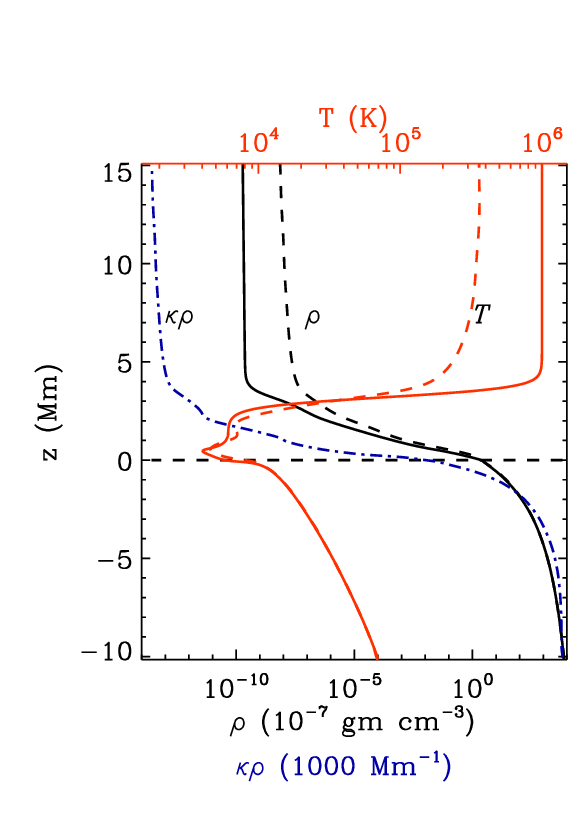}
\put(70,2){\includegraphics[width=0.48\textwidth]{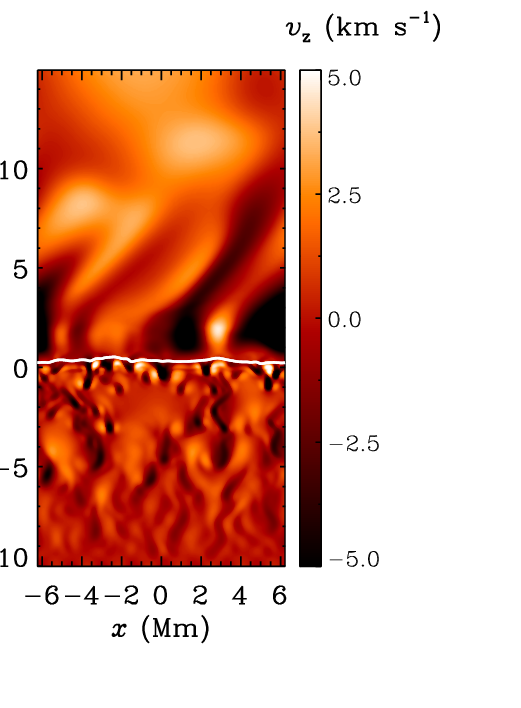}}
\end{overpic}
\caption{\label{fig:profiles} (a) Vertical profiles of the initial (solid) and steady state (dashed) convective temperature (red)and density (black) in the domain. The blue dashed-dotted line shows the profile of $\kappa\rho$ after steady state of convection has been reached. (b) The vertical velocity in the $x-z$ plane at $t=68$ min after the start of the simulation. The white curve shows the location of the optical depth, $\tau=1$ "surface". (color online)}%
\end{figure}
We include explicit height-dependent viscosity, $\nu/\nu_0=1+f(1+\tanh \left\{ (z-z_{1})/w \right\})$, with, $f=4\times10^5$, $z_{1}= 1$ Mm, $w=1.5$ Mm, $\nu_0= 10^{8}$ cm$^2$ s$^{-1}$, whereas for the magnetic diffusivity, we use a uniform and constant value of $\eta_0=10^8$ cm$^2$ s$^{-1}$ unless specified (see table~\ref{tab:setofruns}). The turbulent diffusion, $\chi_t = 10^{8}$ cm$^2$ s$^{-1}$ for $z < 0$ and goes to zero above that. Additionally, we use hyper-diffusion and up-winding to reign in the high frequency wiggles. A density diffusion of $4\times 10^{8}$ cm$^2$ s$^{-1}$ is also included for numerical stability.  
\subsection{Semirelativistic Boris correction to Lorentz force}
\label{sec:boris}

In earlier works, MHD simulations have used Lorentz force limiters of the form 
$(1+v^2_A/c_s^2)^{-1}$ as a multiplying factor for the Lorentz force for low plasma $\beta$ atmospheres. However, \cite{MorCally13} found that these traditional limiters reduce the outward Poynting flux of MHD waves considerably. Instead they explore an alternative method of empirically modifying the density and the gravity from a solar like atmosphere which in turn modifies the acoustic cut-off frequency above the surface. The need for having a semi-relativistic correction to the Lorentz force term in the velocity equation, equation~(\ref{eq:NS}), comes from the fact that in non-relativistic plasmas, the Alfv\'en velocity, $v_A=B/\sqrt{\mu_0\rho}$ can become comparable to the speed of light, $c$.   For example, in the solar corona above the active regions, the magnetic field can be as large as $190$ G inside coronal flux rope \citep{CF13} with a plasma density of $\sim 1.8\times10^{-17}$ g cm$^{-3}$, $v_A/c=0.42$. At these speeds of wave propagation, the plasma can no longer be non relativistic and we cannot neglect the displacement current in the induction equation. While the treatment of electrodynamics is relativistic, the velocity equation remains non-relativistic, making this a semi-relativistic correction. \cite{Boris70} found that by retaining the displacement current, the wave speeds are upper bounded by the speed of light. In order to accelerate the convergence of the explicitly numerical schemes by taking larger time strides, he proposed artificially lowering the $c$. The {\sc Pencil Code} solves for the non-conservative form of the velocity equation, because of which, we follow the primitive variable formulation of the semi-relativistic correction from \cite{Gombosi_etal02}. Here, the velocity equation is modified in the following way.
\EQ
\label{eq:nsboris}
\begin{aligned}
\left[{\bf I}+\frac{v_A^2}{c^2}\left({\bf I}-\hat{\vec{b}}\hat{\vec{b}}\right)\right]{\bm \cdot}\frac{\pa\UU}{\pa t}
   =&-(\UU{\bm \cdot}{\bm \nabla})\UU-\frac{{\bm \nabla} p}{\rho}+ g_z \vec{\hat{z}}
      + &\left[\frac{({\bm \nabla}\times\Bv)}{\mu_0\rho} +\frac{({\bm \nabla}\times\Ev)\times\UU}{\mu_0\rho c^2}\right]\times\Bv.
\end{aligned}
\EN
The term multiplying the ${\pa\UU}/{\pa t}$ on the LHS of equation~(\ref{eq:nsboris}) can be thought of as an ``enhanced inertia" matrix \citep{Rempel17} which makes motion perpendicular to the magnetic field increasingly difficult while the motion parallel to the field lines remains purely hydrodynamic. The limit of validity of the semi-relativistic correction 
is $|\UU| \ll c < v_A$. We follow \cite{Rempel17} and use $c^2=\mathrm{max}(c_s^2, 25|\UU|^2_\mathrm{max})$ 
as the artificially limited speed of light, with $c_s$ denoting the speed of sound. However, we do not 
let $c$ fall below a value of $c_\mathrm{min}=300$ km s$^{-1}$ above the photosphere, whereas, in the convective layer $c$ still has the 
value $3\times10^{10}$ cm s$^{-1}$. For the value of $c_\mathrm{min}$ used, we find that the local Alfv\'en speed $> c_\mathrm{min}$ at about 8000 grid points out of a total of 131072 grid points. These grid points typically located at the top of the box, where the plasma density is lowest, are where the Boris correction applies strongly. Now, let us consider the relative importance of the last two terms 
(inside square brackets) in equation~(\ref{eq:nsboris}). 
\EQ
\frac{|({\bm \nabla}\times\Ev)\times\UU|}{c^2|{\bm \nabla}\times\Bv|} \sim \frac{|\UU|^2}{c^2} \ll 1.
\EN
And, hence the last term consisting of the Electric field, $\Ev$, can be neglected. Note that the viscosity terms have not been modified while implementing the Boris correction. Now, the inverse of the ``enhanced inertia" term, after denoting 
$\beta_A^2=v_A^2/c^2$, can be approximated as,

\EQ
\begin{aligned}
\left[{\bf I}+{\beta_A^2}\left({\bf I}-\hat{\vec{b}}\hat{\vec{b}}\right)\right]^{-1}=&\frac{1}{1+\beta_A^2}\left[{\bf I}+\frac{\beta_A^2}{1+\beta_A^2}\hat{\vec{b}}\hat{\vec{b}}\right] \\
=&{\bf I}-\frac{\beta_A^2}{1+\beta_A^2}\left[{\bf I}-\frac{\hat{\vec{b}}\hat{\vec{b}}}{1+\beta_A^2}\right].
\end{aligned}\label{eq:inertia}
\EN
It is worth noting that the expression of the inverted "enhanced inertia" matrix  in equation~(\ref{eq:inertia}) is more accurate than  equation~(52) of \cite{Rempel17}. With this approximation the correction to the forces on the RHS of equation~(\ref{eq:NS}) is given by

\EQ
F_L^\mathrm{corr}=\frac{\beta_A^2}{1+\beta_A^2}\left[{\bf I} -\frac{\hat{\vec{b}}\hat{\vec{b}}}{1+\beta_A^2}\right]\left[\UU{\bm \cdot}{\bm \nabla}\UU+\frac{{\bm \nabla} p}{\rho}- g_z \vec{\hat{z}}-\frac{({\bm \nabla}\times\Bv)\times\Bv}{\mu_0\rho}\right].
\EN
In the {\sc pencil code}, the Boris semi-relativistic correction can be switched on by setting the flag \texttt{lboris\_correction=T} in the name list \texttt{magnetic\_run\_pars} in the file \texttt{run.in}. According to \cite{Gombosi_etal02}, the characteristic speeds 
due to the Boris correction are modified in a non-trivial manner which is expected to relax the time step constraint due to the 
fastest wave mode. The time step calculation has been modified in the code by replacing the Alfven speed, $v_A$ with $v_A/(1+v_A^2/c^2)^{1/2}$ in the calculation of the Courant criteria. In runs $R_0$--$R_2$, where the magnetic field is only 5 G, the time step is governed mainly by the time step imposed by the \texttt{radiation\_ray} module in the code. However, when we increase the magnetic field strength to 50 G, thereby increasing the Alfv\'en speed, we find that the time step imposed by the \texttt{radiation\_ray} and \texttt{magnetic} modules are comparable. For such large Alfv\'en speeds ($\sim 7000$ km s$^{-1}$), the Boris correction also allows our code to be numerically stable for long simulation times when we set $c_\mathrm{min}=3000$ km s$^{-1}$.  The formulation of the Boris correction in the Pencil code has also been used by \cite{WB19} for a 3D simulation set-up of the solar corona. 

\subsection{The hyperbolic heat transport equation}

\label{sec:hyperbolic}
The anisotropic thermal conduction along magnetic field lines increases to very large values at the high temperatures of the solar corona, constraining the time step severely if the conductivity is treated using explicit numerical schemes. 
Some methods for circumventing this difficulty include either treating the conduction term implicitly or 
by using a time sub-stepping scheme. The time sub-stepping scheme is also available for use in the {\sc pencil code} but we have not tested that for this work. The hyperbolic diffusion equation also known as non-Fickian transport equation 
has been used earlier by several authors in the dynamo community: in the context non-locality of the 
mean field electromotive force using the telegraph equation approach \citep{Branden_etal04, HB09, RB12, BC18}. Such schemes have also been used by \cite{Rempel17} and \cite{Fan17} for treating the anisotropic Spitzer conductivity in the solar corona -- the same purpose as here.
This formulation to treat the Spitzer heat conduction is available in the \texttt{heatflux} module of the {\sc pencil code}. There are three different formulations available again by different authors. Here, we have used the formulation using the subroutine 
\texttt{nonadvective\_nonfourier\_spitzer}. Let $q_\mathrm{cond}$ denote the solution of the non-Fickian transport equation like equation~\ref{eq:qcond} and $q_\mathrm{sp}$ denote the expression of the conduction flux according to the Spitzer model. Then,

\EQ
\label{eq:qcond}
\frac{\pa \vec{q}_\mathrm{cond}}{\pa t}= -\frac{{ \vec{q}_\mathrm{cond}}-
\vec{ q}_\mathrm{sp}}{\tau_\mathrm{sp}}+\beta(\mathrm{\mathrm{d}\vec{r}}{\bm \cdot}\nabla)^6 {\vec{q}_\mathrm{cond}},
\EN
where, 
$$\vec{q}_\mathrm{sp}=K_\mathrm{sp}T^{5/2}\hat{\vec{b}}(\hat{\vec{b}}{\bm \cdot}{\bm \nabla} T),$$
and whenever,
$$\chi_\mathrm{sp}=\frac{K_\mathrm{sp} T^{5/2}}{\rho c_V} > f_{sp} c_0\delta x ,$$
we set the Spitzer diffusion coefficient, $\chi_\mathrm{sp}$ to a fraction $f_\mathrm{sp}$ of the electron free streaming limit  
$c_0\delta x $. Here, $c_0$ is the actual speed of light, $\delta x$ is the maximum grid size and, the Spitzer coefficient, 
$K_\mathrm{sp}$, has a value $10^{-6}$ erg cm$^{-1}$ s$^{-1}$ K$^{-7/2}$ and, 
$\hat{\vec{b}}$ is the unit vector in the direction of the magnetic field including the externally imposed one. We take a small value for the fraction, $f_{sp}=0.1$, $\tau_\mathrm{sp}=0.1$ sec, and the numerical hyper diffusion term,
defined using $\mathrm{d}\vec{r}=(\mathrm{d}x, \mathrm{d}y, \mathrm{d}z)$ and
with $\beta=100$ s$^{-1}$ 
 suppresses numerical wiggles. Compared to the explicit time stepping, we gain a factor of $\sim 100$ on the time step by using this method. Our time step varies between $0.6-10$ ms. 

A sample run directory available as part of the {\sc Pencil code} 
distribution is located at 
\verb|samples/2d-tests/SolarAtmosMag+Boris_Corr+Heatflux| and contains the start (run) parameters in the files \texttt{start.in (run.in)}, the initial stratification file, and a reference output file. We encourage users interested in this simulation to start from this run directory. Additionally, the compilation files \verb|src/Makefile.local| provides the list of physics modules used, and \verb|src/cparam.local| sets the grid size as well as the number of cpus used in each coordinate direction.

\section{Results: Alfv\'en waves in the simulation}
We start our simulation with the density and temperature stratification shown in Fig.~\ref{fig:profiles}(a) 
but with a Gaussian noise with a half width of $\sigma= 0.1$ km s$^{-1}$ for the velocity components. For the first 10 min of solar time, while the initial atmosphere is adjusting to the physics in the simulation, we use a velocity damping in the corona. Due to the super-adiabatic unstable stratification between $-10 ~\mathrm{Mm}<z<0$, 
the convection gradually sets in and the convective energy $\langle\rho U^2\rangle$ reaches a steady value by 
$t= 68$ min. The resultant stratification at $t=68$ min and a snapshot of the vertical velocity is shown in the panels (a) and (b) of Fig.~\ref{fig:profiles}. Note that in this figure as the time in the simulation increases, the temperature in the corona part of the domain decreases due to the optically thin cooling profiles used and in absence of any explicit heating function. This cooling causes the density in the corona to also increase. This will be a challenge for any heating source trying to re-heat the collapsing, be it wave dissipation or reconnection, as it has more coronal material to heat in order to increase the temperature. 
The residual magnetic field in the simulation atmosphere is very weak and 
we do not expect the dynamics to be affected due to magnetic reconnection. 
The external field, which is not acted upon by the flow, is oriented in the 
$\hat{\vec{x}}+\hat{\vec{z}}$ direction and is increased from 0.1 G to 5.0 G gradually within a duration of $75$ min 
of solar time. After this $B_\mathrm{ext}$ is kept fixed at $5.0$ G. This gradual increase of $B_\mathrm{ext}$ is necessary since this prevents the numerical value of density from reducing to very low values during the adjustment of the initial atmosphere to the included physics.  During this time the maximum Alfv\'en speed in the domain increases from 210 km s$^{-1}$ to 770 km s$^{-1}$. The $B_\mathrm{ext}$ inclined at an angle $\theta=45^o$ to the vertical is used to facilitate the conversion of fast magneto-acoustic modes to Alfv\'en modes as suggested by \cite{CG08} as well as \cite{CH11} where they found that the conversion efficiency depends sensitively on angles $\theta$ and the azimuth, $\phi$ at which the fast waves are incident on the external magnetic field. In our case $\phi=0$ for convective motions and $\phi=90^0$ for the motion excited due to the forcing $\mathcal{F}_y$ described below, where as \cite{CH11} find that $\phi\sim60^0-90^0$ and $\theta\sim 30^0-40^0$ leads to near total conversion of fast mode to Alfv\'en waves and is completely absent for $\theta=0$ or for purely vertical magnetic fields.
 
Since this is a 2.5 dimensional setup, we expect all the MHD modes -- slow and fast magnetoacoustic waves, 
linearly and circularly polarized Alfv\'en waves can be excited in the simulation.  As the magnetic field in the domain is homogenous, we do not expect surface modes like torsional or sausage modes in this simulation. These surface modes constitute another vast field of study and have also been detected in the upper solar chromosphere \citep{Jess_etal09} using data from the Swedish Solar Telescope. However, here our aim is to only study the propagation 
and damping properties of shear Alfv\'en waves excited due to the convection and in the context of the semi relativistic Boris correction implemented here. We find that the amplitude of $U_y$ 
is quite less ($\sim 0.1$ km $s^{-1}$) because of the 2.5 dimensional rather than full 3 dimensional set up.  
We separate the velocity vector into components in normal, $\hat{\vec{n}}=(\hat{\vec{z}}-\hat{\vec{ x}})/\sqrt{2}$ 
and tangential, $\hat{\vec{t}}=(\hat{\vec{z}}+\hat{\vec{x}})/\sqrt{2}$ directions with reference to the external 
magnetic field which points towards $\hat{\vec{t}}$.  So, in order to drive a linearly polarized Alfv\'en wave with 
perturbations in the $\hat{\vec{y}}$ direction, we force the system between $0.6<z<1$ Mm with a forcing 
$\mathcal{F}_y\sim u^{ff}_{y} \sin(k_x x+k_z z-\omega t)/\omega$ in the $y$-component of the velocity equation. Note that the driving is only in terms of the perturbed velocity and not the magnetic field $\vec{B}'$. We expect that the induction equation will produce a perturbed magnetic field corresponding to this velocity driving. We consider the time period 
of the driving to be $2\pi/\omega=100$ sec and amplitude $u^{ff}_{y}=1$ km s$^{-1}$. Note that this driving introduced in the interior of the box is different from that at the bottom boundary. We additionally require this internal driving as the strength of the velocity driving at the bottom boundary by itself is unable to generate enough power in the variability of the $y$-component of the velocity. However, we start this 
driving only for $t > 83$ min after the start of the simulation, long after the external magnetic field has reached its 
final value of 5.0 G.  For now, the external magnetic field is taken to be uniform in contrast to that expected in the solar atmosphere where the magnetic field would exist in flux tubes or loops. The top boundary does reflect some of the MHD waves as we haven't imposed the characteristic equations for the magneto-acoustic and Alfv\'en waves at the top. But, this is alright as wave reflection is a common phenomena in the line-tied magnetic loops in the atmosphere.

\begin{figure}
\begin{overpic}[width=0.33\textwidth]{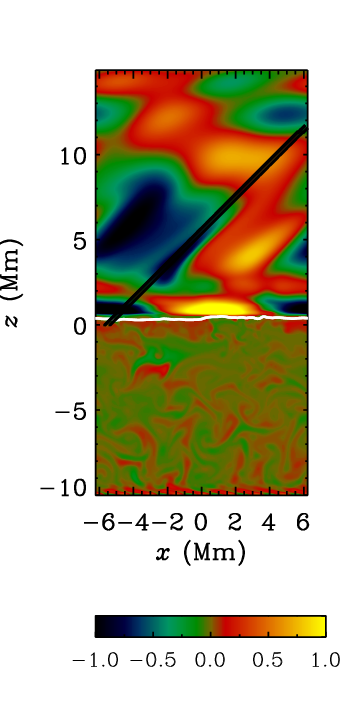}
\put(20,95){(a) $U_y (\mathrm{km~s}^{-1})$}
\end{overpic}
\begin{overpic}[width=0.33\textwidth]{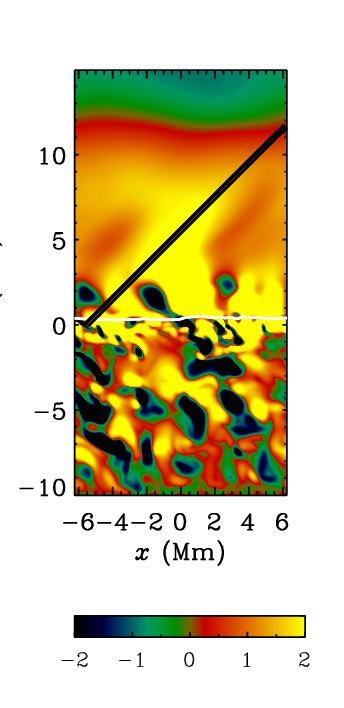}
\put(10,95){(b) $U_n=\frac{U_z-U_x}{\sqrt{2}} (\mathrm{km~s}^{-1})$}
\end{overpic}
\begin{overpic}[width=0.33\textwidth]{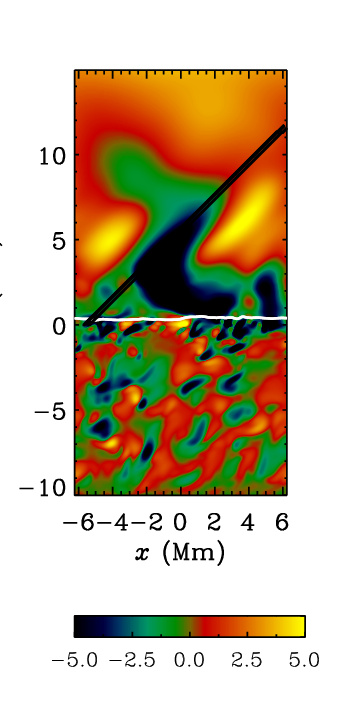}
\put(10,95){(c) $U_t=\frac{U_z+U_x}{\sqrt{2}} (\mathrm{km~s}^{-1})$}
\end{overpic}
\caption{\label{fig:vyvnvt} Snapshots of velocity components in $\hat{\vec{y}}$, $\hat{\vec{n}}$ and, $\hat{\vec{t}}$ directions at a time $t=100.5$ min after the start of the simulation run, $R_0$. The black slanted lines denote the location of the slit (along the $\hat{\bm t}$-direction) while the white curve is the optical depth, $\tau=1$ surface. This figure is available as an animation. (color online)}
\end{figure}

\begin{figure}
\begin{overpic}[width=0.33\textwidth]{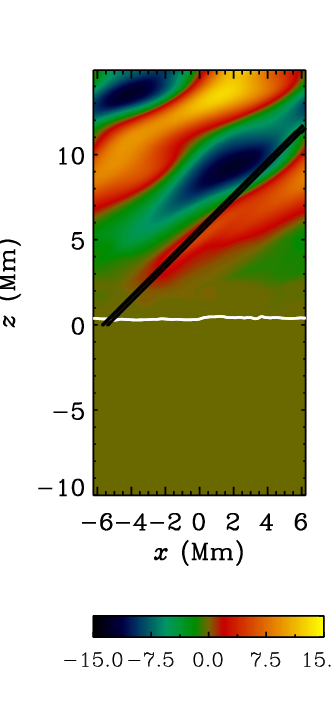}
\put(20,95){(a) $v_{Ay}' (\mathrm{km~s}^{-1})$}
\end{overpic}
\begin{overpic}[width=0.33\textwidth]{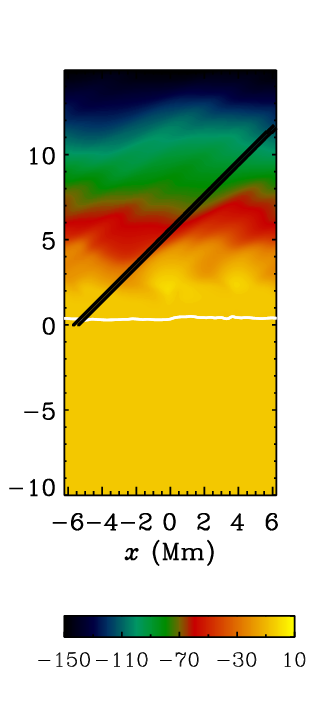}
\put(10,95){(b) $v_{An}'=\frac{v_{Az}'-v_{Ax}'}{\sqrt{2}} (\mathrm{km~s}^{-1})$}
\end{overpic}
\begin{overpic}[width=0.33\textwidth]{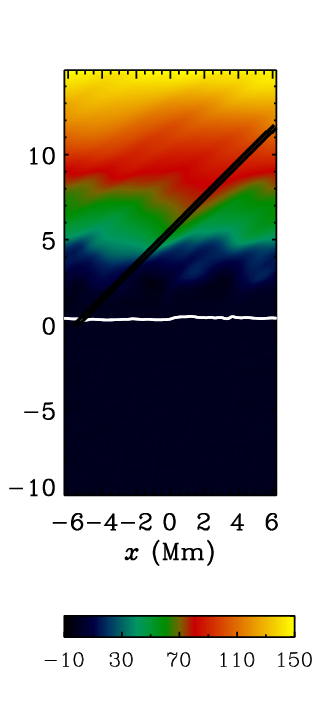}
\put(10,95){(c) $v_{At}'=\frac{v_{Az}'+v_{Ax}'}{\sqrt{2}} (\mathrm{km~s}^{-1})$}
\end{overpic}
\caption{\label{fig:bybnbt} Snapshots of components of Alfv\'en velocity, $\vec{v}_A'$, in $\hat{\vec{y}}$, $\hat{\vec{n}}$ and, $\hat{\vec{t}}$ directions at a time $t=100$ min after the start of the simulation for run $R_0$. The black slanted lines denote the location of the slit (along the $\hat{\bm t}$-direction) and the white curve is the optical depth, $\tau=1$ surface. This figure is available as an animation. (color online)}
\end{figure}
\begin{figure}
\begin{overpic}[width=0.95\textwidth]{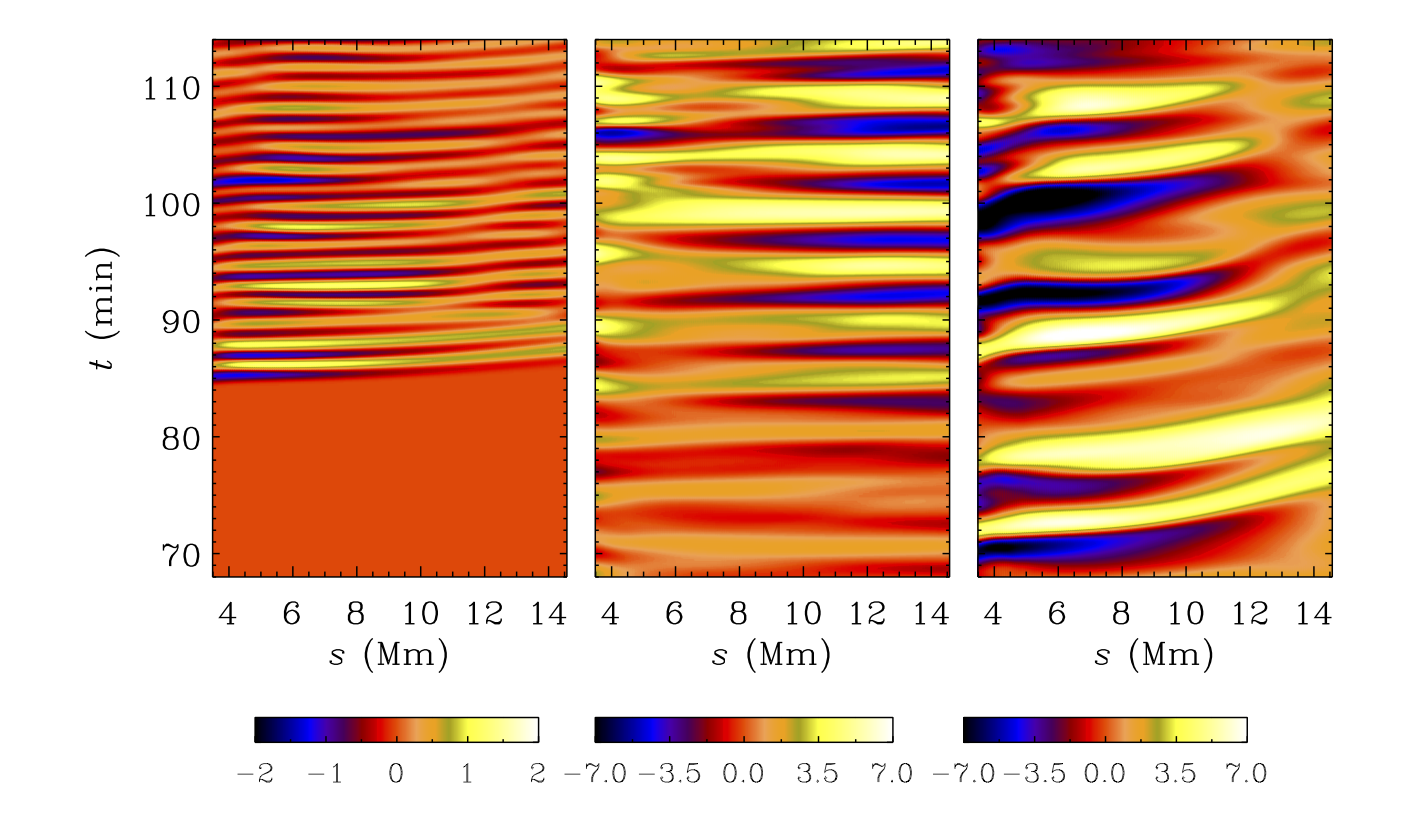}
\put(20,57){(a) $U_y$}
\put(45,57){(b) $U_n=\frac{U_z-U_x}{\sqrt{2}}$}
\put(70,57){(c) $U_t=\frac{U_z+U_x}{\sqrt{2}}$}
\end{overpic}
\caption{\label{fig:velynt} The time-distance maps for run $R_0$ corresponding to the simulation duration between $70<t<112$ min averaged over the width of the slit shown in the panels of figure~\ref{fig:vyvnvt} for (a) $U_y$, (b) $U_n$ and (c) $U_t$. The abcissa is the length along the slit and the velocity components are in units of km s$^{-1}$. (color online)}
\end{figure}
\begin{figure}
\begin{overpic}[width=0.95\textwidth]{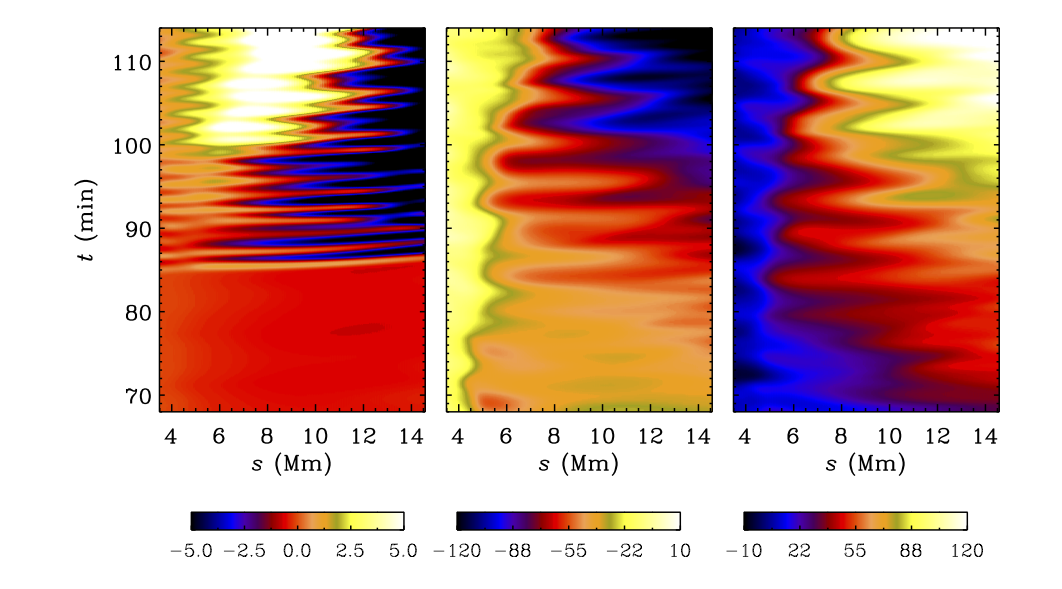}
\put(20,55){(a) $v_{Ay}'$}
\put(45,55){(b) $v_{An}'=\frac{v_{Az}'-v_{Ax}'}{\sqrt{2}}$}
\put(70,55){(c) $v_{At}'=\frac{v_{Az}'+v_{Ax}'}{\sqrt{2}}$}
\end{overpic}
\caption{\label{fig:bbynt} The time-distance maps for the run $R_0$ corresponding to the simulation duration between $70<t<112$ min averaged over the width of the slit shown in the panels of figure~\ref{fig:vyvnvt} for (a) $v_{Ay}'$, (b) $v_{An}'$ and (c) $v_{At}'$. The abscissa is the length along the slit.  (color online)}
\end{figure}

\begin{center}
\begin{table}
 \caption{\label{tab:setofruns} Summary of the runs. The parameters are defined in the text. The run $R_0$ is the original run described in detail whereas the runs, $R_1$, $R_2$ and $R_3$ and branched from a particular snapshot of 
 $R_0$ at different times.}
 \centering
  \begin{tabular}{@{}lrrrr@{}}
  \\
    \toprule
  & {\bf $R_0$} & {\bf $R_1$} & {\bf $R_2$} & {\bf $R_3$}\\
    \midrule
{\bf $f$}  & $4\times10^5$ & $4\times10^5$& $4\times10^3$&$4\times10^4$\\
$\eta_0$ (cm$^2$ s$^{-1}$)& $10^8$ & $10^8$& ${2\times10^{11}}$&$8\times10^{12}$\\
$Pr_M$ & $8\times10^5$ & $8\times 10^5$ & $4$&$1$\\
 $k_xL_x/2\pi$& $1$ & $1$&$1$ & $8$\\
$u^{ff}_y$(km s$^{-1}$) & $1$& $5$& $5$&$2$ \\
    \bottomrule
  \end{tabular}
 \end{table}
\end{center}
The snapshots of the three different components of the velocity - $U_y$, $U_n$, and $U_t$ - at a time $t=100.5$ min for run, $R_0$, is shown in figure~\ref{fig:vyvnvt}. Panel (a) of this figure shows propagation of positive and negative $U_y$ phases along the direction of the external guide field. The snapshot in panels (b) have clear signatures of modes that are reflected where as, panel (c) which shows $U_t$ shows patterns aligned in the {\bf $t$} direction. From the animation corresponding to this figure, we clearly see waves propagating in the $+x$ direction. The snapshots of the perturbation $\vec{v}_A'$ in figure~\ref{fig:bybnbt} is more complex. For example, the contours of constant phase are mis-aligned with the external magnetic field. For Alfv\'en waves, we expect $\UU$ and $\vec{v}_A'$ to have equal amplitudes so as to satisfy equipartition of energy between magnetic field and the  velocity. This seems not to be the case here. The likely reasons for this is the large magnetic Prandtl number ($Pr_M=\nu/\eta \gg 1$) as well as modification of the structure of the modes due to the semi relativistic correction. 
We present  time evolution diagrams 
for the velocity and magnetic field components in figure~\ref{fig:velynt} and figure~\ref{fig:bbynt} along a slit shown by the black line in Figures~\ref{fig:vyvnvt}, \ref{fig:bybnbt} aligned along the $\hat{\bm t}$-direction (same as the direction of the externally applied magnetic field). The slit width over which the pattern has been averaged is 0.3 Mm.  In panel (a) of Figures \ref{fig:velynt} and \ref{fig:bbynt}, the amplitudes of $U_y$ and $v_{Ay}'$ are substantial and has a time period of 1.67 min only after $t=83$ min which is a clear signature of the driving $\mathcal{F}_y$.
In both figure~\ref{fig:velynt} and figure~\ref{fig:bbynt}, the first thing we notice is that the time-period for 
$U_y$ and $v_{Ay}'$ (1.67 min from panel a) is less than that of other components in the $\hat{\vec{n}}$ and $\hat{\vec{t}}$ directions (4.8 min and 5.5 respectively in panels b and c). The time periods have been calculated from the inspection of the waveforms in regions where the amplitudes are stronger, {\it e.g.}, at $s=4$ Mm for components $U_y$ and $U_t$ and at $s=12$ Mm for the component $U_n$. For $t< 100$ min, we have only forced the velocity in the $y$-direction, exciting a linearly polarized Alfv\'en wave mode. Also, the presence of non-zero velocity component in the $\hat{\vec{t}}$ direction (along $\vec{B}_\mathrm{ext}$) implies excitation of 
magneto-acoustic waves, as Alfv\'en waves are transverse and cannot have a velocity component in the direction of the external guide field.  
In fact, this component, $U_t$, has a consistent amplitude throughout the simulation duration and is not 
affected by the forcing (in the y-direction) switched on for $t>83$ min and is a result of the 2-dimensional convection. The average speeds of propagation is fastest for $U_n$ in panel 
(b) and slowest for $U_t$ in panel (c) of  figure~\ref{fig:velynt}.

\begin{figure}
\begin{overpic}[width=0.9\textwidth]{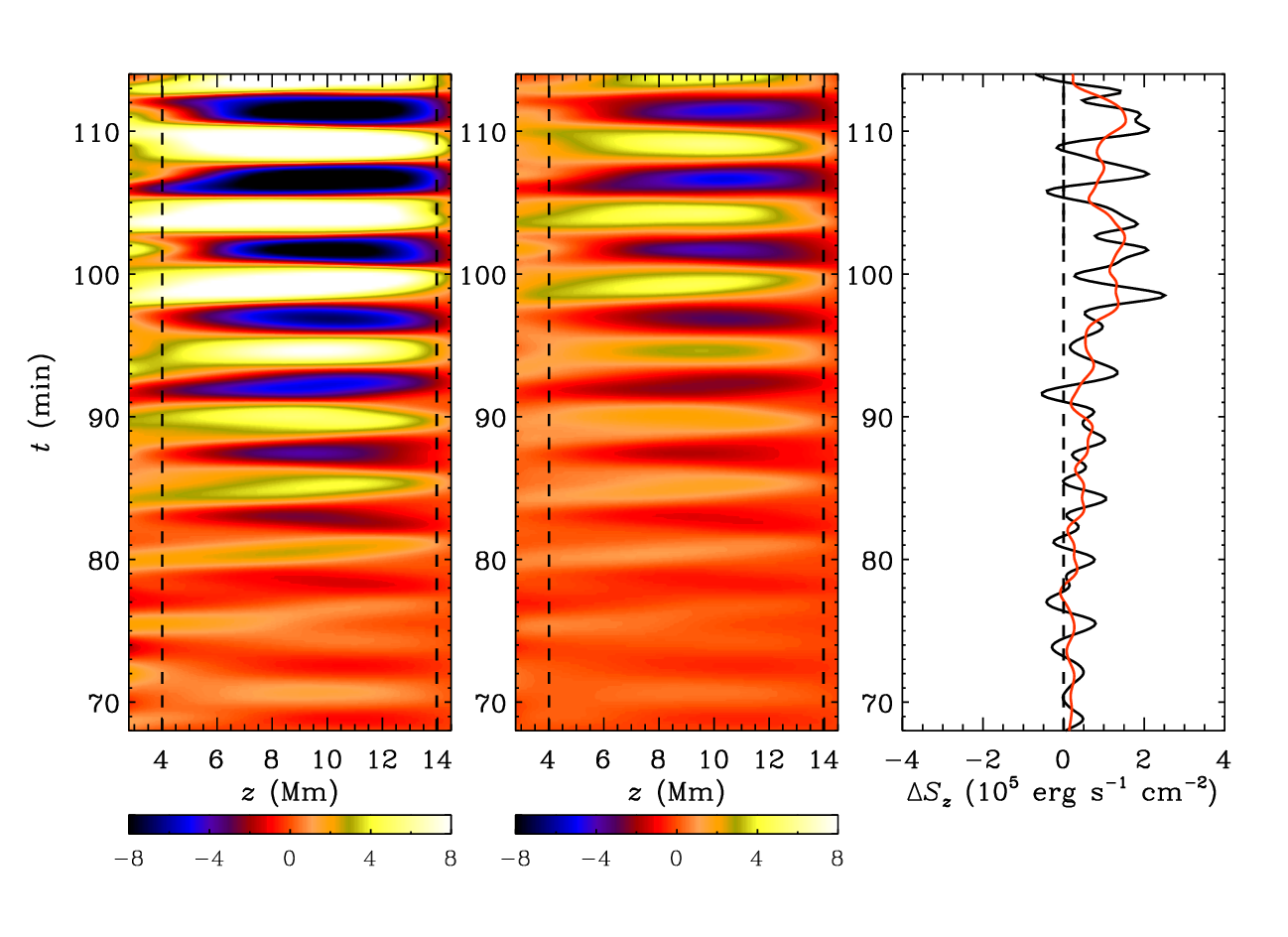}
\put(12,70){\small (a) $S_z$}
\put(42,70){\small (b) $-\langle\rho(\vec{v}_{A\perp}'{\bm \cdot}\UU_\perp)v_{Az}\rangle$}
\put(75,70){\small (c) $T_\mathrm{max}$ (MK)}
\put(12,3){\small $10^7$ ergs cm$^{-2}$ s$^{-1}$}
\put(42,3){\small $10^7$ ergs cm$^{-2}$ s$^{-1}$}
\end{overpic}
\caption{\label{fig:pfz} (a) The vertical component of the horizontally averaged Poynting flux as a function of time for $R_0$. The time period $\sim 5$ min. (b) The contribution to the vertical Poynting flux from transverse waves. (c) $\Delta S_z = S_{z1}-S_{z2}$ (black), calculated by taking $z_1=4$ Mm and $z_2=13.9$ Mm denoted by dashed lines in panel (a). The red curve is the running mean calculated with a time interval 2.5 min. (color online)}
\end{figure}

We show the $z$-component of the horizontally averaged Poynting flux, $\vec{S}$,  only due to the waves in figure~\ref{fig:pfz}. We calculate this flux using
$$S=\frac{1}{L_x}\int\frac{\vec{B}'\times(\UU\times\Bv)}{\mu_0} \mathrm{d}x .$$
The vertical component $S_z$ shows an increase of amplitude when the velocity forcing is switched on.  The dominant time period as clear from panel (a) is 4.8 min which is the same periodicity as seen in $U_n$ in figure~\ref{fig:velynt}(b). The 
alternating sign of negative and positive Poynting flux is a signature of modes interfering due to reflection. 
If the wave energy existed only as transverse waves, for which $\vec{B}'{\bm \cdot}\Bv=0$, the Poynting flux term will be $\propto -\rho (\vec{v}_{A}'{\bm \cdot}\UU)\vec{v}_A$. A better way to estimate the contribution of the transverse Alfv\'en waves to $S_z$ would be to 
estimate $-\langle\rho(\vec{v}_{A\perp}'{\bm \cdot}\UU_\perp)v_{Az}\rangle$, where $\langle \rangle$ denotes 
horizontal averaging and, with $\vec{v}_{A\perp}'={v}_{An}'\hat{\vec{n}}+v_{Ay}'\hat{\vec{y}}$ (see panel b of figure~\ref{fig:pfz}). Clearly, the transverse waves have a major contribution to the spatio-temporal pattern of  $S_z$. In panel (c) of figure~\ref{fig:pfz}, we show the net Poynting flux injected in the region $4~\mathrm{Mm}< z < 13.9~\mathrm{Mm}$ by the simple formula $\Delta S_z=S_{z1}-S_{z2}$ (solid black line). When 
$\Delta S_z >0$ for extended time intervals, excess energy is available for heating the corona, where as for $\Delta S_z <0$ also for extended time intervals, the corona between $z_1$ and $z_2$ will cool during this time interval. The running average of $\Delta S_z$ over a time interval 
2.5 min is shown in red. We see that average of $\Delta S_z$ can reach values of $10^5$ ergs cm$^{-2}$ s$^{-1}$ in the lower solar corona below 15 Mm.  This value can be compared with a Poynting flux of $5\times 10^7$ ergs s$^{-1}$ cm$^{-2}$ produced by the photospheric driver corresponding to an active region (without any well formed sunspot pair observed by the SOT instrument on board the Hinode satellite on 14 Nov, 2007) used in a 3D MHD model for coronal loops \citep{Bourdin_etal2014, Bourdin_etal2015}. 
We also performed an alternate run $R_1$, branching from the original run $R_0$ presented above at $t=98$ min, where we only increased the amplitude, $u^{ff}_y$, of the forcing, $\mathcal{F}_y$, by a factor of five. Now, these waves cannot be classified as small amplitude waves anymore. Interestingly, we found that in this run, the amplitude 
$U_y $ is often a  factor of about 10 smaller than $-v_{Ay}'$ (see figure~\ref{fig:vyby}) at $t=103.8$ min.  As we have mentioned earlier that in the semi relativistic limit the Alfv\'en speeds are quite different from the classical MHD case and that may explain the factor of 10 difference in the fluid velocity and Alfv\'en velocity perturbations. However, the patterns of $U_y$ and $v_{Ay}'$ again diverged 
after $t=110$ min. This may have happened due to the very large explicit magnetic Prandtl number chosen in the corona part of 
the simulation, inspired by the expression for viscosity and resistivity given by Spitzer (1962). We tested this hypothesis 
successfully by reducing $Pr_M \sim 4$ (i.e., by reducing $f=10^3$) in yet another run $R_2$ where $u^{ff}_y=0.5$ and 
branched from the run $R_0$ at $t=83$ min and the patterns of $U_y$ and $v_{Ay}'$ continued to be in anti-phase for 
20 min after the restart from $R_0$. For run $R_2$, the amplitude $U_y$ is again a factor of ten smaller than $-v_{Ay}'$. 
From panels (a) and (b) of this figure we can infer that the plane of polarization of these propagating waves lies in a direction given by $\vec{k}\times\vec{B}_\mathrm{ext}$ (or in the $\hat{\vec{y}}$ direction) 
where, $\vec{k}$ is the wave vector and makes an angle $\sim 15^{o}$ with the guide field, $\vec{B}_\mathrm{ext}$. 
Also we can see that the magnitude of the wave vector is $\sim 2\pi/Lx$. The dispersion relation for incompressible 
Alfv\'en waves in a homogenous atmosphere without stratification can be easily obtained by linearizing the MHD equations~\ref{eq:NS}, ~\ref{eq:induc} (without the Boris correction term) and is given by, 
\EQ
\label{eq:disp}
\omega^2+i(\nu+\eta)k^2\omega-\nu \eta k^4-v^{2}_Ak^2=0.
\EN
This is a very simple relation and is not correct for our present simulation set-up. Nevertheless, the imaginary part of the solution of this quadratic equation is given by $i(\nu+\eta)k^2/2$. An approximate estimate for decay time of Alfv\'en waves in our simulation using the above equation is $\sim 33$ min which is will lead to a very slow heat rate by wave dissipation. One way of reducing this decay time will be to use a large wavenumber in the $x$-direction for the driving $\mathcal{F}_y$ \citep{RH17}.  We also test this by increasing the horizontal wave number in the run, $R_3$.  The real part of the frequency, from the dispersion relation in equation~(\ref{eq:disp}), has the form $\sqrt{v_A^2k^2-(\nu-\eta)^2k^4}/2$. For large $Pr_M$, there would be a region in the corona where  Alfv\'en waves (for small Alfv\'en velocities and large $k$ like here) will be evanescent. However, for $Pr_M=1$ this problem does not arise. Therefore, in the run $R_3$, we also change the Prandtl number to be one.  Besides, we also change the height for the forcing to lie between 
$2~\mathrm{Mm}<z<2.2~\mathrm{Mm}$, so that the forcing frequency $\omega$ can take a value of $v_A k \sim 0.02\pi$ rad s$^{-1}$. The waves produced by the driving in $R_3$ propagate in a direction more aligned to the field $\vec{B}_\mathrm{ext}$ (not shown) in comparison to the run $R_0$ (see Figs.~\ref{fig:vyvnvt}(a) and~\ref{fig:vyby}(a)). We started this run from an 
early snapshot of run $R_0$ at $t=76.3$ min. The decay time for the $kL_x/2\pi=8$ is expected to be 0.05 min from the simple equation~(\ref{eq:disp}). Nevertheless, we see that the maximum temperature in the domain (during the simulation duration of 
53 min of solar time after introduction of the forcing at $t=76.3$ min) increases and fluctuates between $2\times10^5< T<1.2\times10^6$ presumably due to Alfv\'en wave dissipation as opposed to the run $R_0$ (panel c of figure~\ref{fig:ts}). It is worth noting 
that for all the other runs, the temperature in the corona part of the domain decreases with time logarithmically rather than 
exponentially as seen in the panel (c) of figure~\ref{fig:ts} (solid black line).  Note that, the forcing $u_y^{ff}$ was not switched 
on at the time when the runs--$R_1$, $R_2$ and $R_3$--branched from $R_0$. It is possible that shock fronts appearing in the simulation run, $R_3$, as a result of branching from run, $R_0$, by means of change of parameters like the Prandtl number has 
caused these localized temperature enhancements. But, this is unlikely as we found episodes where the maximum temperature 
exceeded 1 MK in a run with same parameters like $R_3$ but initiated at $t=0$ from same initial conditions as $R_0$ instead 
of branching from it. In this test run, after the episodic bursts of high temperature, the maximum temperature eventually 
decayed like the other runs, $R_0$, $R_1$ and $R_2$. Because of the above reasons, we do not claim this heating as a consistent feature arising out of Alfv\'en wave dissipation. 
The Poynting flux is larger in magnitude in the run $R_3$ as compared to the run $R_0$ as shown in figure~\ref{fig:ts}(a). Until $t=120$ min, the time-distance diagram of the vertical Poynting flux show similar slopes for positive and negative phases in both Figs.~\ref{fig:pfz}(a) and \ref{fig:ts}(a) but, after this the speed of propagation seems to decrease for the run $R_3$. This may be because the propagation direction becomes more horizontal and the effective vertical speed decreases as evident from animation of the snapshots of $U_y$, $U_n$ and, $U_t$ in the domain available as an animation file: run3.mp4 in accompanying supplementary material. In the animation the white arrows denote the direction (not magnitude) of the total magnetic field including $\vec{B}_\mathrm{ext}$. We should also note that in run $R3$, the waves excited may be finite amplitude waves which can modify the guide magnetic field from the given $\vec{B}_\mathrm{ext}$.
In the panel (b) of the same figure, we show the contribution of the transverse waves to the Poynting flux, 
namely, $\langle-\rho(\vec{v}_{A\perp}'{\bm \cdot}\UU_{\perp})\vec{v}_A\rangle$. Here also like in figure~\ref{fig:pfz}(b), the transverse waves majorly contribute to the spatio-temporal pattern of  $S_z$. 

\begin{figure}
\begin{overpic}[width=0.3\textwidth]{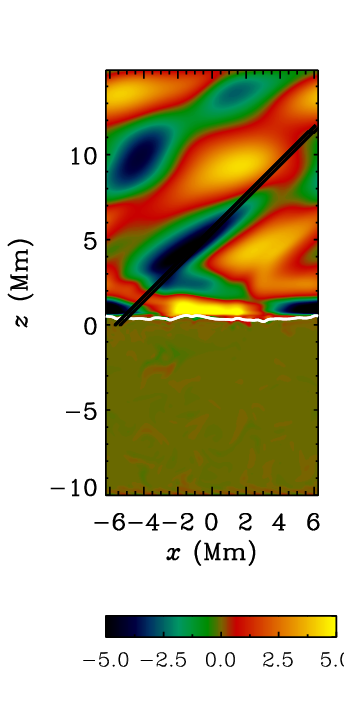}
\put(20,95){(a) $U_y$}
\end{overpic}
\begin{overpic}[width=0.3\textwidth]{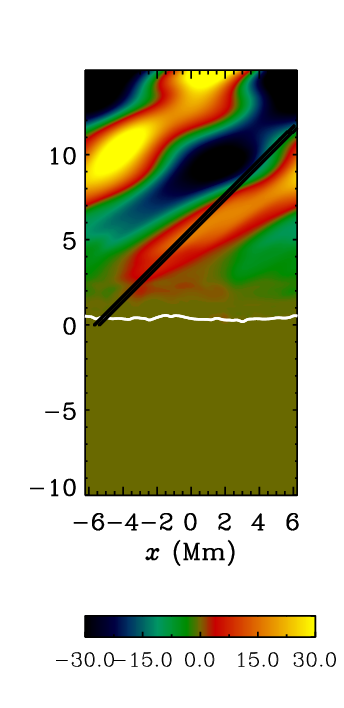}
\put(20,95){(b) $v_{Ay}'$}
\put(40,14){\includegraphics[width=0.4\textwidth,height=0.52\textwidth]{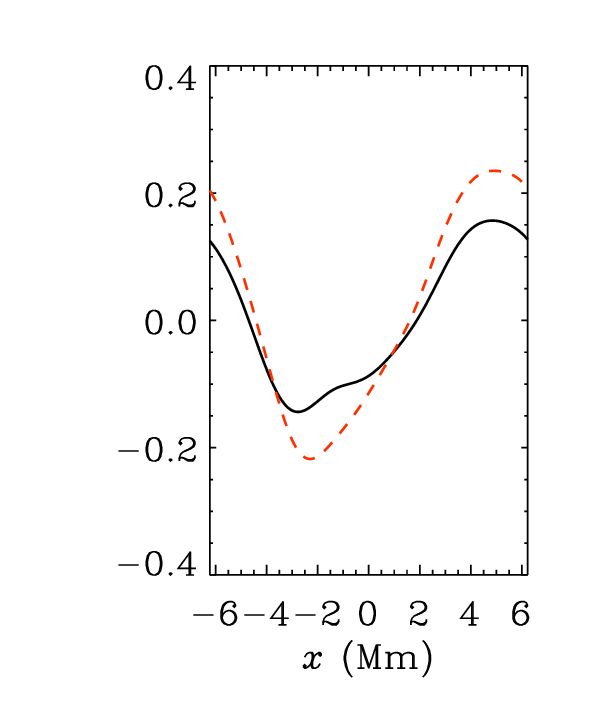}}
\put(75,95){(c)}
\end{overpic}
\caption{\label{fig:vyby} (a) A snapshot of $U_y$ in the domain at $t=103.8$ min for the run $R_1$ with the forcing enhanced after $t>98$ min. (b) A snapshot of $v_{Ay}'$ at the same time. (c) Horizontal profiles of $U_y$ (solid) and 
$-v_{Ay}'/10$ (dashed) at $z=12$ Mm. (color online) }
\end{figure}
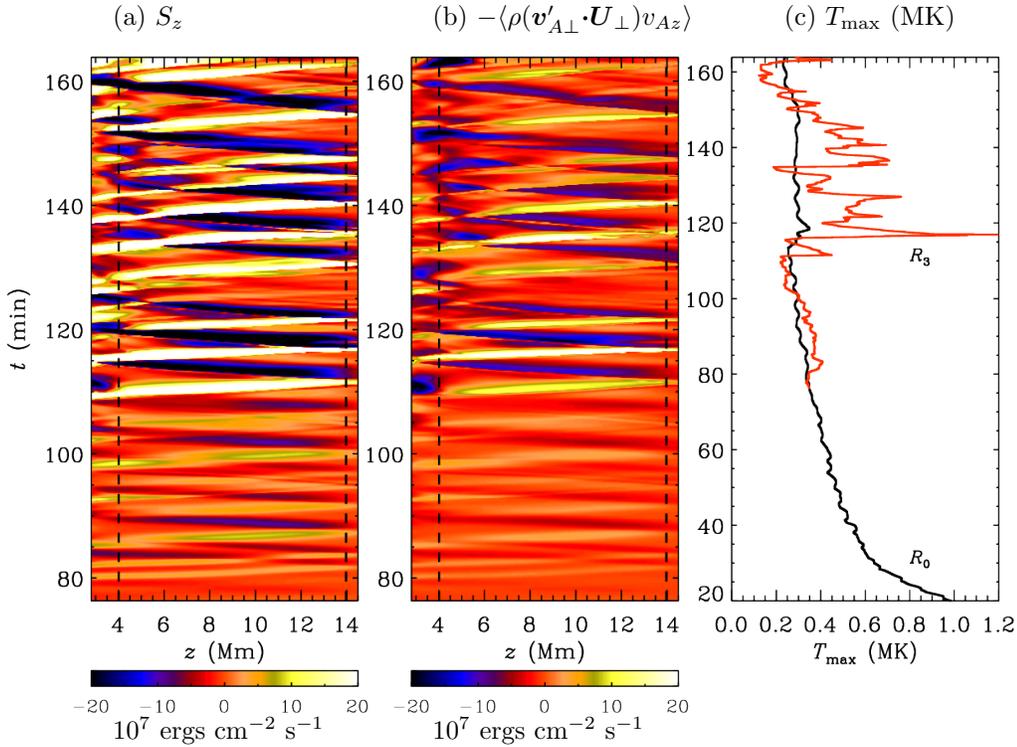
\begin{figure}
\begin{overpic}[width=0.8\textwidth]{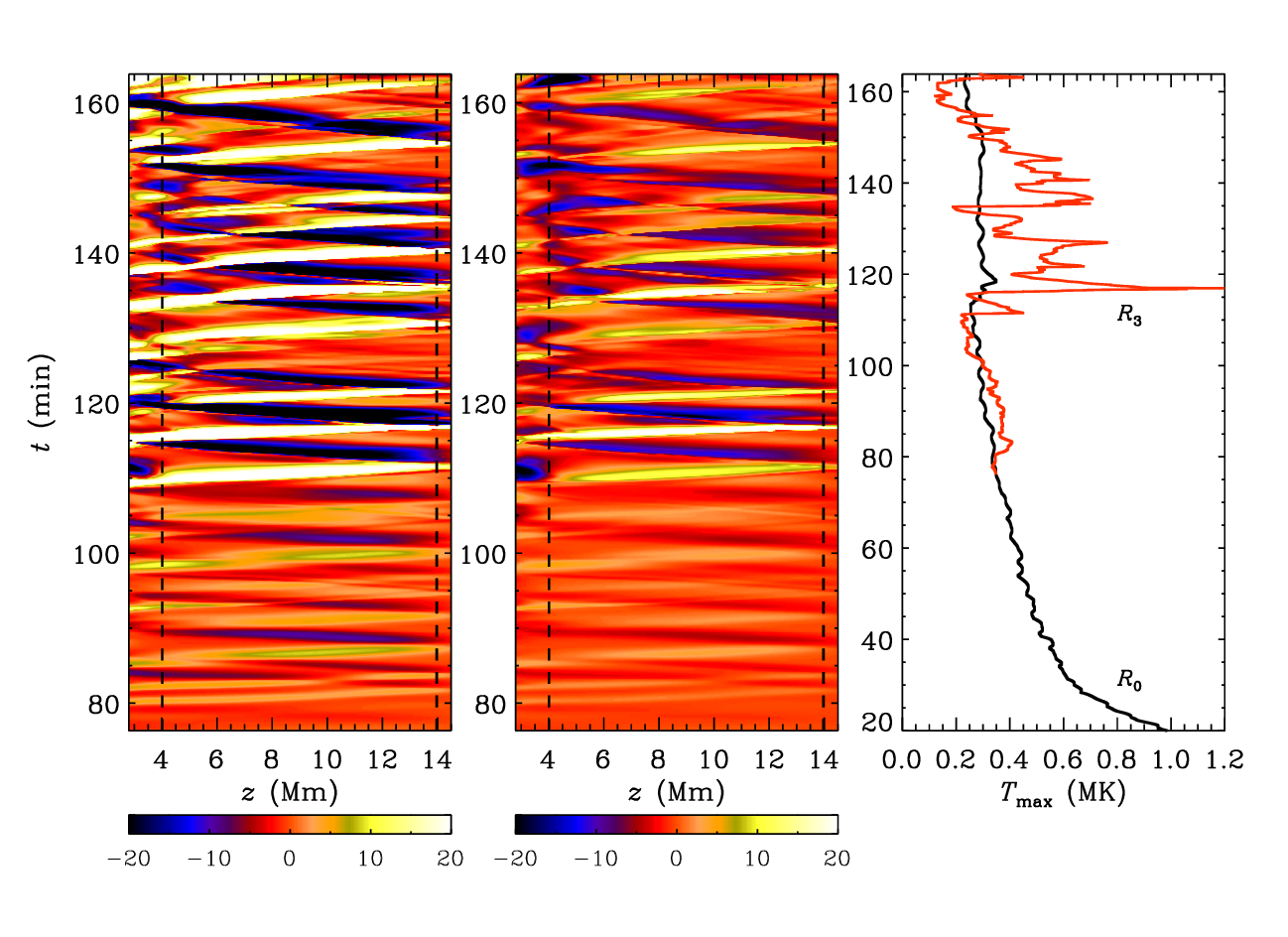}
\put(12,70){\small (a) $S_z$}
\put(42,70){\small (b) $-\langle\rho(\vec{v}_{A\perp}'{\bm \cdot}\UU_\perp)v_{Az}\rangle$}
\put(75,70){\small (c) $T_\mathrm{max}$ (MK)}
\put(12,3){\small $10^7$ ergs cm$^{-2}$ s$^{-1}$}
\put(42,3){\small $10^7$ ergs cm$^{-2}$ s$^{-1}$}
\end{overpic}
\caption{\label{fig:ts} (a) The vertical component of the horizontally averaged Poynting flux as a function of time for $R_3$. (b) The contribution to the vertical Poynting flux from transverse waves. (c) The time-evolution of the maximum temperature in the domain for the two runs $R_0$ (black solid) and $R_3$ (red). (color online)}
\end{figure}
\section{Discussion}
We present a 2.5 dimensional Cartesian radiative MHD setup using the {\sc pencil code} in a 
first step towards understanding 
the propagation of Alfv\'en waves in an atmosphere with a ``realistic" solar like stratification. We incorporated 
the semi relativistic Boris correction in the velocity equation for numerical stability and also use a hyperbolic heat transport equation to bypass the time stepping constraint due to an explicit field aligned thermal conduction. 
We consider spatial variations only in the $x-z$ plane with $z$ as the direction of stratification. 
The velocity and the magnetic field has components in the $y$-direction 
but has no variation with the $y$ axis. The lower part of the box is convective below $z<0$, but being 2.5 dimensional the amplitude of $U_y$ is much smaller than $U_x$ and $U_z$. Hence, in order to excite Alfv\'en waves, we drive the system with an artificial forcing, $\mathcal{F}_y$, in the $y$-component of the velocity equation with a frequency 100 Hz and an amplitude such that the mean $U_y$ between $0.6~\mathrm{Mm}<z<1$ Mm is $\sim 0.7$ km s$^{-1}$. This value is still a factor of five smaller than the other velocity components in the $x-z$ plane. The forcing $\mathcal{F}_y$ can excite linearly polarized Alfv\'en waves with the plane of polarization perpendicular to the vector 
$\hat{\vec{n}}$ which can propagate out of the domain with a group 
velocity $\vec{v}_A$. This is evident from the fact that the time taken to traverse a distance of 10 Mm along the slit is $\sim 1$ min from figure~\ref{fig:velynt} (a) and matches with the estimate obtained by integrating $\int \mathrm{d}s/v_A(s)$ between $4~\mathrm{Mm} < z< 14~\mathrm{Mm}$, where $ds$ is the infinitesimal element along the slit and $v_A(s)$ is the Alfv\'en speed along the external guide field. We believe that the mismatch in the spatial and temporal patterns 
of fluid velocity and the perturbation magnetic field, $\Bv'$ in the run $R_0$ is due to a very large value of the 
Prandtl number used in the coronal part of the simulation.  The runs here have been continued for about 110 min of solar time during which the dynamics have been influenced by the magnetic Prandtl number as well as the nature of the 
forcing employed. For all runs reported in table~\ref{tab:setofruns} except $R_3$, the temperature in the corona 
decreases logarithmically.  For $R_3$, we see some heating in form of local enhancements of temperature which 
is intermittent in time but still decreases during the end of simulation run. The external magnetic field used in 
this set-up is not the correct reproduction of the magnetic field topologies in the Sun where Alfv\'en 
velocities will be large in the chromosphere and transition region coronal loops. So far, we have tested and found the 
set-up to be stable for maximum Alfv\'en speeds reaching as large as $4.04$ Mm s$^{-1}$. In the future simulations 
we wish to explore external magnetic field resembling coronal loop-like geometries and 
strengths rather than a simple inclined uniform and weak magnetic field used here. The driving 
we have used here is located much above the photosphere. With coronal loop like external magnetic fields we can circumvent 
the need of this artificial driving at $\sim 2$ Mm as the driving velocities will be channelled to the 
corona along the strong magnetic loops from underneath the photosphere much faster. In this 2.5 dimensional set-up, however we cannot do away with the forcing $\mathcal{F}_y$ completely since $|U_y| < |U_{x,z}|$, but, the forcing can then be set up underneath the photosphere rather than above it. It is also important to explore the role of reconnection which we have neglected here and compare with the contributions of wave heating. These are the topics of future explorations using the {\sc Pencil code} set-up described here.  

\section*{Acknowledgements}
We thank the two anonymous referees for carefully reviewing the manuscript which has lead to considerable improvement upon its clarity. We acknowledge the computing time awarded on PARAM Yuva-II supercomputer at C-DAC, India under the grant name Hydromagnetic-Turbulence-PR. Most of the simulation runs were carried out on the HPC cluster, Nova, of the Indian Institute of Astrophysics.  We thank P. A. Bourdin and J\"orn Warnecke for discussions on the code especially related to the solar corona module. We also thank Sahel Dey, for his project work on dispersion relations of Alfv\'en waves.

\newcommand{\yana}[5]{, #5.  {\em Astron.\ Astrophys.\ }{#1}, {\bf #2}, #3-#4.}
\newcommand{\yanaS}[5]{, #5. {\em Astron.\ Astrophys.\ }{#1}, {\bf #2}, #3-#4.}
\newcommand{\yapj}[5]{, #5. {\em Astrophys.\ J.\ }{#1}, {\bf #2}, #3-#4.}
\newcommand{\yjgr}[5]{, #5. {\em  J.\ Geophys.\ Res. }{#1}, {\bf #2}, #3-#4.}
\newcommand{\yapjss}[5]{, #5. {\em Astrophys.\ J.\ Supp. Series }{#1}, {\bf #2}, #3-#4.}
\newcommand{\yjfm}[5]{, #5. {\em J.\ Fluid Mech.\ }{#1}, {\bf #2}, #3-#4.}
\newcommand{\yjcp}[5]{, #5. {\em J.\ Com. Phy.\ }{#1}, {\bf #2}, #3-#4.}
\newcommand{\ymn}[5]{, #5. {\em Mon.~Not.~ R.\ Astron.\ Soc.\ }{#1}, {\bf #2}, #3-#4.}
\newcommand{\ysph}[5]{, #5. {\em Solar Phys.\ }{#1}, {\bf #2}, #3-#4.}
\newcommand{\ypre}[5]{, #5. {\em Phys.\ Rev.\ E }{#1}, {\bf #2}, #3-#4.}
\newcommand{\yprlN}[5]{, #4.  {\em Phys.\ Rev.\ Lett. }{#1}, {\bf #2}, #3.}
\newcommand{\yjourN}[5]{, #5. {\em #2} {\bf #3}, #4.}
\newcommand{\yza}[5]{, #5. {\em Zeitschrift f\"ur Astrophysik } {#1}, {\bf #2}, #3-#4.}
\newcommand{\ypasj}[5]{, #5. {\em Publications of the Astronomical Society of Japan }{#1}, {\bf #2}, #3-#4.}

\end{document}